\documentclass[12pt]{article}
\begin{document}

\begin{center}
A Novel "Magnetic" Field And Its Dual Non-Commutative Phase Space
\vskip 2cm

Subir Ghosh\\
Physics and Applied Mathematics Unit,\\
Indian Statistical Institute,\\
203 B. T. Road, Calcutta 700108, India.
\end{center}
\vskip 3cm
{\bf Abstract:}\\
In this paper we have studied a new form of Non-Commutative (NC)
phase space with an operatorial form of noncommutativity. A point
particle in this space feels the effect of an  interaction with an
"{\it{internal}}" magnetic field, that is singular at a specific
position $\theta^{-1}$. By "internal" we mean that the effective
magnetic fields depends essentially on the particle properties and modifies the symplectic structure.
Here $\theta $ is the NC parameter and induces the coupling
between the particle and the "internal" magnetic field. The
magnetic moment of the particle is computed.  Interaction with an
{\it{external}} physical magnetic field reveals interesting
features induced by the inherent fuzziness of the NC phase space:
introduction of non-trivial structures into the  charge and  mass  of the particle and  possibility of the particle dynamics collapsing
to a Hall type of motion. The dynamics is studied both from
Lagrangian and symplectic (Hamiltonian) points of view. The
canonical (Darboux) variables are also identified. We briefly comment, that the model presented here, can
play interesting role in the context of (recently observed)
{\it{real}} space Berry curvature in material systems.

\newpage
Particle motion in an electromagnetic field and its  {\it{dual}}
description in terms of an effective Non-Commutative (NC)
(phase)space is very well known. However, with the  advent of
various forms of {\it{intrinsically}} NC phase spaces
\cite{sw,sn,kappa} in recent times, the above scenario has turned
in to an actively discussed topic. To give the simplest of
examples, the Lorentz force law,
\begin{equation}
m\ddot x_i=e\epsilon_{ijk}\dot x_jB_k, \label{01}
\end{equation}
is derivable from either of the two Hamiltonian systems given
below:
\begin{equation}
\{x_i,p_j\}=\delta_{ij}~;~\{x_i,x_j\}=\{p_i,p_j\}=0~;~~H=\frac{1}{2m}(p_i-eA_i)^2,
\label{02}
\end{equation}
and
\begin{equation}
\{x_i,P_j\}=\delta_{ij}~;~\{P_i,P_j\}=e\epsilon_{ijk}B_k~;~~\{x_i,x_j\}=0~;~~H=\frac{1}{2m}(P_i)^2,
\label{03}
\end{equation}
 where time evolution is given by $\dot O=\{O,H\}$. The equivalence is complete if one invokes validity of the Jacobi identity{\footnote{In our classical setup, commutators are replaced by Poisson Brackets.}} in the latter algebra,
$$\epsilon_{ijk}\{p_i,\{p_j,p_k\}\}\sim \partial_{i}B_i=0,$$ which demands source free magnetic fields with  $B_i=\epsilon_{ijk}\partial_{j}A_k$.  The   two systems are related  by $P_i\equiv p_i-eA_i$. However, we emphasize that this identification is not necessary and both symplectic structures (\ref{02}) and (\ref{03}) are each of them can be utilised independently. On the other hand, in quantum mechanics, associativity - the key property - can be maintained even with $\partial_{i}B_i\ne 0$ but the source has to be monopole like \cite{jac1}.

More close to the framework of NC geometry is the Landau problem that deals with a point charge $e$ in a plane with a constant magnetic  field $B$ normal to the plane. In the limit of large $B$, the Lagrangian and the corresponding symplectic structure reduces to,
\begin{equation}
L=-\frac{B}{2}\epsilon_{ij}\dot x_ix_j~\Rightarrow \{x_i,x_j\}=\frac{\epsilon_{ij}}{B},
\label{04}
\end{equation}
which is the NC plane. Although the above are  not examples of an intrinsically NC phase space, the same mechanism will work in an intrinsically NC phase space, that can be simulated by an "internal" magnetic field. This is the subject matter of the present paper.

It is quite obvious, at least mathematically, that an NC phase
space, complimentary to (\ref{03}), is possible  where the
coordinates are noncommutative, (instead of the momenta), which
will yield a monopole in momentum space. An explicit construction
of such a spacetime first appeared in $2+1$-dimensional spinning
particle models \cite{cnp} of Anyons \cite{wil} - excitations with
arbitrary spin and statistics. Very interestingly, singular
structures of precisely the above monopole form have been
experimentally observed \cite{mono1} in {\it{momentum}} space. The
theoretical description relies on the momentum space Berry
curvature and associated monopole singularity in anomalous Hall
effect \cite{niu}. This is closely connected to the exciting areas
of spin Hall effect and {\it{spintronics}} \cite{spin}. The
condensed matter phenomena and NC spacetime were first connected
in a coherent framework by Berard and Mohrbach \cite{berard} who
have considered an NC coordinate space with a singularity in
momentum variable. This is in fact complimentary to the NC
momentum space of \cite{jac1} mentioned above, that describes a
charged particle in the presence of a magnetic monopole in
coordinate space. The idea of \cite{berard} has been further
elucidated in \cite{duv} who demonstrate that Bloch electron
dynamics, developed in \cite{niu} is indeed Hamiltonian in nature,
 provided the proper
symplectic phase space volume is taken into account. This
clarifies the observations earlier made in \cite{xiao}, (see also
\cite{xiao1}), that in this case the {\it{naive}} phase space
volume does not remain conserved, due to the presence of Berry
curvature and magnetic field.

This has naturally given rise to the question of existence of
singular structures in coordinate space and a recent paper
\cite{pur} has suggested observation of the same. The theoretical
framework \cite{gen} is based on the non-vanishing local spin
chirality.

In this perspective we put forward the motivations of our work.
Our primary aim is to construct an NC phase space that induces an
effective singular magnetic field in {\it{coordinate}} space, that
is structurally different from a monopole. We claim that this
result can be relevant in the observation of coordinate space
singularity in  \cite{pur}.  The NC phase space we have presented
here is new and has not appeared before. As we will elaborate, it
is complimentary to the well known NC spacetime model proposed by
Snyder \cite{sn}. For this reason we refer to our construction as
the "Snyder" space and since we develop a point particle picture,
we term the latter as a "Snyder" particle. The other aim is a
generalization of the work of \cite{berard}, which is  restrictive
in the sense that the noncommutativity depends {\it{only}} on
momentum variables and this is due to the fact that in
\cite{berard} the mixed coordinate-momentum bracket is taken as
canonical: $\{x_i,p_j\}=\delta_{ij}$ {\footnote{We have shown in
\cite{ban} that  more general forms of NC phase space yields
structurally different forms of singular magnetic fields, in
momentum space.}}. In our work,  Berry phase in coordinate space
appear. We emphasize that our generalizations (here and in
\cite{ban}) {\it{maintains rotational invariance}} in the sense
that the angular momentum operator and algebra as well as
rotational transformation laws of position and momentum are
unchanged. This is explained later in terms of Darboux variables.
However, the singular nature  of the intrinsic magnetic field is
{\it{not}} of the form of a monopole.  We deal with a
non-relativistic classical model but the major part of our work
goes through in the quantized version.

After studying dynamics of the non-interacting "Snyder" particle
we couple it with an {\it{external}} physical electromagnetic
field. Now the inherent fuzziness of the NC phase space is
revealed whereby the localized properties of the particle, its
charge and mass, aquire  non-trivial structures. In fact, we find
a quadrupole moment like behavior in the particle charge
distribution which is interesting if we recall \cite{jabb} that
fundamental particles behave like dipoles in spaces with constant
noncommutativity.

Let us start by considering a (first order) Lagrangian,
\begin{equation}
L=\dot X_iP_i+\theta\dot X_i{\cal A}_i -H~,~~{\cal A}_i=\frac{
(X.P)}{1-\theta X^2}X_i , \label{7}
\end{equation}
where $\theta $ denotes the particle charge, $H$ is some
unspecified Hamiltonian and the vector potential ${\cal A}_i$ is
given by,
\begin{equation}
{\cal A}_i=\frac{ (X.P)}{1-\theta X^2}X_i . \label{7a}
\end{equation}
We have used the notation $(X.P)=X_iP_i,~X^2=X_iX_i$. This leads
to a magnetic field of the form,
\begin{equation}
{\cal B}_i=\epsilon_{ijk}\partial^{(X)}_{j}{\cal A}_k =
\epsilon_{ijk}\frac{P_jX_k}{1-\theta X^2}\equiv
-\frac{L_i}{1-\theta X^2}, \label{8}
\end{equation}
where $L_i=\epsilon_{ijk}X_jP_k$ is the particle angular
momentum{\footnote{We will establish later that $L_i$ is truly the
conserved angular momentum.}}. It might be possible to express $\vec {\cal B}$ as a
combination of multipole fields. Notice that $\partial_{i}{\cal
B}_i=0$ but for the singular surface $ X^2=\theta^{-1}$

Let us consider a (magneto)static property of the particle,
{\it{i.e.}} its magnetic moment. Mimicking the classical
electrodynamics laws for magnetostatics, for a steady current
${\cal J}_i$ one has ${\cal J}_i=\epsilon_{ijk}\partial_{j}{\cal
B}_{k}$, which leads to an effective magnetic moment,
\begin{equation}
{\cal M}_i=\frac{1}{2}\epsilon_{ijk}X_j{\cal
J}_k=\frac{1}{(1-\theta X^2)^2}L_i. \label{mm}
\end{equation}
Quite interestingly, even for this "Snyder" particle residing in
an NC space, (to be elaborated later), the above  agrees with the
conventional result for a moving particle \cite{jack}.

Before studying the dynamics, let us see the effect of introducing
the particular form of "internal" magnetic field. It is termed
internal because it actually reflects the property of the particle
in question, rather than a physical external magnetic field. In
fact it modifies the symplectic structure, as we now demonstrate.

For a generic first order Lagrangian, as in (\ref{7}),
\begin{equation}
L=a_{\alpha}(\eta )\dot \eta^{\alpha}-H(\eta ), \label{3a}
\end{equation}
with $\eta$ being phase space variables, one has the
Euler-Lagrange equations of motion,
\begin{equation}
\omega_{\alpha\beta}\dot \eta^{\beta}=\partial_{\alpha}H~
\Rightarrow \dot
\eta^{\alpha}=\omega^{\alpha\beta}\partial^{\beta}H. \label{3b}
\end{equation}
In the above, $\omega_{\alpha\beta}$ denotes the symplectic two
form \cite{fj}. In the present model (\ref{7}) we thus compute,
\begin{equation}
\omega_{\alpha\beta}=
 \left (
\begin{array}{cc}
-\frac{\theta}{1-\theta X^2}(X_iP_j-X_jP_i)  & -(\delta_{ij}+\frac{\theta}{1-\theta X^2}X_iX_j) \\
(\delta_{ij}+\frac{\theta}{1-\theta X^2}X_iX_j) &  0
\end{array}
\right ) . \label{4}
\end{equation}

 We adhere to our earlier notation (\ref{3a})
\cite{fj,duv}, that defines the symplectic structure for a first
order Lagrangian model in the following way:
\begin{equation}
\{f,g\}=\omega^{\alpha\beta}\partial_{\alpha}f\partial_{\beta}g.
\label{6}
\end{equation}
In the above, $f$ and $g$ constitute two generic operators and
$\omega^{\alpha\beta}$ is the inverse of the symplectic matrix,
\begin{equation}
\omega^{\alpha\beta}\omega_{\beta\gamma}=\delta^{\alpha}_{\gamma}~,~~
\omega_{\alpha\beta}=\partial_{\alpha}a_{\beta}-\partial_{\beta}a_{\alpha}.
\label{3}
\end{equation}
For the model (\ref{7}) we compute,
\begin{equation}
\omega^{\alpha\beta}=
 \left (
\begin{array}{cc}
 0 & (\delta_{ij}-\theta X_iX_j) \\
-(\delta_{ij}-\theta X_iX_j) &  -\theta (X_iP_j-X_jP_i)
\end{array}
\right ) . \label{5}
\end{equation}
It induces the non-canonical Poisson Brackets in a straightforward
way:
\begin{equation}
\{X_i,X_j\}=0~,~~\{X_i,P_j\}=\delta_{ij}-\theta
X_iX_j~,~~\{P_i,P_j\}=-\theta (X_iP_j-X_jP_i). \label{1}
\end{equation}
This is the new form of  NC space, or  "Snyder" space, that we
have advertised and this constitutes the other major result in our
paper. For $\theta =0$ we recover the canonical phase space. This
can be compared with the  spatial part of Snyder algebra \cite{sn}
which reads,
\begin{equation}
\{X_i,X_j\}=-\theta
(X_iP_j-X_jP_i)~,~~\{X_i,P_j\}=\delta_{ij}-\theta
P_iP_j~,~~\{P_i,P_j\}=0, \label{1a}
\end{equation}
and the effects of which have been studied in various contexts
\cite{chang}. The duality between (\ref{1}) and (\ref{1a}) is due to the fact we had in mind the construction of an NC phase space that would be rotationally invariant and have a singular structure in the effecetive magnetic field in {\it{coordinate}} space. From our previous experience with Snyder algebra in \cite{ban}, we knew that Snyder phase space does have this property but with the singularity in {\it{momentum}} space. Thus, we guessed, (which turns out to be  correct), that a form of NC spacetime will be needed that is dual to the Snyder form.

Notice that the structures (\ref{1}) (and (\ref{1a}) as well)
survive upon quantization since there are no operator ordering
ambiguities. {\footnote{Actually, the $\{P_i,P_j\}$ bracket
requires a symmetrization but the effect cancels out due to
anti-symmetry.}} More important for us is the fact that these NC
algebras satisfy the Jacobi identity,
\begin{equation}
J(A,B,C)\equiv  \{\{A,B\},C\}+\{\{B,C\},A\}+\{\{C,A\},B\} =0,
\label{jac}
\end{equation}
with $A,B$ and $C$ being $X_i$ or $P_j$.

Indeed, from the viewpoint of symplectic analysis, it is not
surprising that the induced phase space (\ref{1}) is associative
(in the sense of classical brackets). However, at the same time we
should remember that in the Lagrangian framework (\ref{7}) of the
same system, we had a particle interacting with a singular
magnetic field and singular potentials tend to violate Jacobi
identities \cite{jac1,berard}. But there is no controversy
involved{\footnote{I thank Peter Horvathy for discussions on this
point.}}  since, although the final result (\ref{1}) has no
ambiguity regarding associativity, the intermediate steps
($\omega_{\alpha\beta}$ or its inverse) are defined  only for
$X^2\neq \theta^{-1}$.

We now consider the dynamics in the  Faddeev-Jackiw symplectic
formalism \cite{fj} (see also \cite{duv}). We consider a simple
form of Hamiltonian
\begin{equation}
H=\frac{ P^2}{2m} , \label{h}
\end{equation}
in analogy with the normal non-interacting particle Hamiltonian
and
 the particle dynamics follows:
\begin{equation}
\dot X_i=\frac{i}{m}(\delta_{ij}-X_iX_j)P_j~\Rightarrow
P_i=m(\delta_{ij}+\theta \frac{X_iX_j}{1-\theta X^2})\dot X_j,
\label{eq1}
\end{equation}
\begin{equation}
\dot P_i=\frac{\theta}{m}((X.P)P_i-P^2X_i) \Rightarrow ~m\ddot
X_i=-m\theta [(2-\theta X^2)\dot X^2+\frac{\theta}{1-\theta
X^2}(X.\dot X)^2]X_i. \label{eq2}
\end{equation}

It is worthwhile to obtain the rotation generator. We posit it to
be of the form
\begin{equation}
L_i=\epsilon_{ijk}X_jP_k . \label{c2}
\end{equation}
We find that it transforms the position $X_i$ and momentum $P_i$
properly:
\begin{equation}
\{L_i,X_j\}=\epsilon_{ijk}X_k~,~~\{L_i,P_j\}=\epsilon_{ijk}P_k .
\label{c3}
\end{equation}
It satisfies the correct angular momentum algebra and is
conserved,
\begin{equation}
\{L_i,L_j\}=\epsilon_{ijk}L_k ~;~~ \{L_i,H\}=0. \label{c4}
\end{equation}
This demonstrates the consistency of our derivation of the
magnetic moment of the particle in (\ref{mm}).

Since, as shown in (\ref{mm}), there is an effective magnetic
moment involved even with the non-interacting particle, it is more
interesting to study the behavior of the "Snyder" particle in the
presence of external electromagnetic field. This is done by
introducing the external $U(1)$ gauge field $C_i$ in (\ref{7}),
\begin{equation}
a_i=\theta {\cal {A}}_i+ eC_i. \label{cc4}
\end{equation}
We thus obtain,
\begin{equation}
\omega_{\alpha\beta}^{(e)}=
 \left (
\begin{array}{cc}
-\frac{\theta}{1-\theta X^2}(X_iP_j-X_jP_i)+eF_{ij}  & -(\delta_{ij}+\frac{\theta}{1-\theta X^2}X_iX_j) \\
(\delta_{ij}+\frac{\theta}{1-\theta X^2}X_iX_j) &  0
\end{array}
\right ) . \label{4e}
\end{equation}
and its inverse,
\begin{equation}
\omega^{(e)\alpha\beta}=
 \left (
\begin{array}{cc}
 0 & (\delta_{ij}-\theta X_iX_j) \\
-(\delta_{ij}-\theta X_iX_j) &  -\theta
(X_iP_j-X_jP_i)+e[F_{ij}+\theta (F_{ki}X_j-F_{kj}X_i)]X_k
\end{array}
\right ) . \label{5e}
\end{equation}
where $F_{ij}=\partial_{i}C_{j}-\partial_{j}C_{i}$ and
$\omega^{(e)}$ signifies the presence of external interaction.
This modifies the original symplectic structure to:
$$
\{X_i,X_j\}=0~;~~\{X_i,P_j\}=\delta_{ij}-\theta X_iX_j~,$$
\begin{equation}
\{P_i,P_j\}=-\theta (X_iP_j-X_jP_i) +e[F_{ij}+\theta
(X_iF_{jk}-X_jF_{ik})X_k] \label{6e}
\end{equation}
The new equations of motion are,
\begin{equation}
\dot X_i=(\delta_{ij}-\theta X_iX_j)\frac{P_j}{m}, \label{e4}
\end{equation}

\begin{equation}
\dot
P_i=e(E_i+\frac{1}{m}F_{ij}P_j)+\frac{\theta}{m}[(X.P)P_i-P^2X_i]+\theta
e[\frac{1}{m}(X.P)X_kF_{ki}-\frac{1}{m}X_kP_jF_{kj}X_i-(X.E)X_i],
\label{e5}
\end{equation}
where the Hamiltonian is taken as,
\begin{equation}
H=\frac{P^2}{2m}+e\phi (X). \label{e6}
\end{equation}
The equation of motion for $X_i$ throws up some interesting
features which are due to the NC phase space with its inherent
fuzziness.
\begin{equation}
m\ddot X_i=e[\delta_{ij}-\theta (2-\theta
X^2)X_iX_j](E_j+F_{jk}\dot X_k)-m\theta [(2-\theta X^2)\dot
X^2+\frac{\theta}{1-\theta X^2}(X.\dot X)^2]X_i. \label{e7}
\end{equation}
The Lorentz force term in the R.H.S. shows that we are no longer
dealing with a simple point charge but an effective charge tensor that
reminds us of a quadrupole moment structure, which, in the normal
spacetime is expressed as,
\begin{equation}
Q_{ij}=\int (3x'_ix'_j-r'^2\delta_{ij})\rho (\vec
x')d^3x'=3X_iX_j-X^2\delta_{ij},
 \label{e9}
\end{equation}
where the last relation is for a localized distribution. One
should recall \cite{jabb} that an NC spacetime with a constant
form of noncommutativity also induces a dipole like behavior in
the basic excitations.

In an earlier interesting work, Duval and Horvathy \cite{dh} had
considered an "exotic" particle in the two-dimensional plane, that
is characterized by {\it{two}} parameters, the mass and another
scalar parameter, the latter being related to the Anyon spin in
the non-relativistic limit of \cite{cnp}. The work \cite{dh}
showed that, when placed in a constant magnetic field normal to
the plane, (the classical Landau problem), the particle aquires a
dressed mass and the condition of vanishing of this effective mass
forces the particle to follow the Hall law. It might be
interesting to see whether a similar situation prevails here. With
this motivation we recast (\ref{e7}) in the form,
\begin{equation}
m(\delta_{ij}+\theta \frac{2-\theta X^2}{(1-\theta
X^2)^2}X_iX_j)\ddot X_j=e(E_i+F_{ij}\dot
X_j)-\frac{m\theta}{(1-\theta X^2)^2}[(2-\theta X^2)\dot
X^2+\theta \frac{(X.\dot X)^2}{1-\theta X^2}]X_i. \label{e8}
\end{equation}
Once again we note that there is an effective tensorial structure  instead of a
scalar point mass. Let us simplify (\ref{e8}) by considering $O(\theta )$
effects only and in the mass-tensor we make a spatial averaging
$\frac{X_iX_j}{X^2}=3\delta_{ij}$ and obtain,
\begin{equation}
m^*\ddot X_i=e(E_i+F_{ij}\dot X_j)-2\theta m^* X_i+ O(\theta ^2).
 \label{e13}
\end{equation}
In the above, the effective mass $m^*$ is given by,
\begin{equation}
m^*=m(1+6\theta X^2). \label{e14}
\end{equation}
Taking cue from \cite{dh}, we conclude that for a negative value
of $\theta$, $m^*$ can vanish on the surface $x^2\sim (6\theta
)^{-1}$ in which case we find the Hall law of motion,
\begin{equation}
E_i+F_{ij}\dot X_j ~\Rightarrow \dot
X_i=\frac{1}{B}\epsilon_{ij}E_j, \label{e15}
\end{equation}
where for the planar case $F_{ij}=B\epsilon_{ij}$.

Rotational transformation of the phase space variables reveals
that, with $L^{ij}=X^iP^j-X^jP^i$, the coordinates behave as
before,
\begin{equation}
\{L_{ij},X_k\}=\delta_{ik}X_j-\delta_{jk}X_i. \label{ee9}
\end{equation}
As expected, the momentum transformation is modified by the gauge
field term but there is an additional $e\theta $-contribution
which appears due to the "Snyder" NC phase space but vanishes for
$e=0$. This is qualitatively different from the previously studied
result (\ref{c3}) with no $U(1)$ interaction
\begin{equation}
\{L_{ij},P_k\}=\delta_{ik}P_j-\delta_{jk}P_i
+e(X_iF_{jk}-X_jF_{ik})-e\theta (X_iF_{jl}-X_jF_{il})X_lX_k.
\label{e10}
\end{equation}

We will  end our discussion on the particle dynamics scenario with
a few words on canonical (Darboux) coordinates corresponding to
the phase space (\ref{1}) of non-interacting "Snyder" particle. It
is easy to check that the algebra (\ref{1}) can be reproduced by
expressing the NC variables $(X_i,P_j)$ in terms of a canonical
set $(x_i,p_j)$ obeying
$$\{x_i,p_j\}=\delta_{ij}~, \{x_i,x_j\}=\{p_i,p_j\}=0~,$$
via the transformations:
\begin{equation}
X_i\equiv x_i~,~~P_i\equiv p_i-\theta (x.p)x_i~. \label{11}
\end{equation}
Quite obviously, these transformations will require proper
operator ordering upon quantization. Exploiting (\ref{11}) we
find,
\begin{equation}
L_{ij}=X_iP_j-X_jP_i=x_ip_j-x_jp_i,
 \label{l}
\end{equation}
which shows that the angular momentum operator remains
structurally unchanged. In terms of the $(x,p)$ set, the
Lagrangian (\ref{7}) will reduce to,
\begin{equation}
L=\dot x_ip_i -H(x,p). \label{12}
\end{equation}
The symplectic matrix has turned in to a canonical one as expected
and the interactions will be manifested through the Hamiltonian.
From the structure of the Hamiltonian in $(x_i,p_j)$ canonical
space, it might seem that the singular behavior of the "Snyder"
particle, discussed throughout this paper, has disappeared.
However, we reemphasize that the inverse of the above Darboux
transformation (\ref{11}) exists provided $X^2\neq \theta^{-1}$.

Finally, we come to the works \cite{duv}, (related to
\cite{berard}),  that have discussed the condensed matter problem
of Berry phase effect in the context of anomalous Hall effect
\cite{niu,mono1} from a fluid dynamical framework in phase space.
It has been shown in \cite{duv} that the modified equations of
motion \cite{niu} for a Bloch electron, in the presence of a Berry
curvature, are indeed Hamiltonian in nature provided one exploits
an NC phase space and uses the appropriate (phase space)
volume-form. One can attempt similar analysis in our "Snyder" NC
phase space (\ref{1}), which is quite distinct from the ones
considered in \cite{berard,duv}. In the present case, the correct
volume form \cite{duv} is,
\begin{equation}
\sqrt{det \omega_{\alpha\beta}}=\frac{1}{1-\theta X^2}. \label{ll}
\end{equation}
Notice that the same singularity has reappeared in the volume
element.

Let us comment on the possible physical realization of our model.
A very recent paper \cite{pur} has suggested the possible
observation of real space Berry phase structure in the anomalous
Hall effect in re-entrant AuFe alloys. As has been pointed out in
\cite{gen,pur}, this real space Berry phase contribution depends
on topologically nontrivial contributions of spin and involves the
magnetization explicitly. Hence clearly the structure of the
singularity will be more complicated than a monopole type and will
have a chiral nature. It should also be remembered that
\cite{gen1} in principle, complicated structures of the vortex are
indeed possible depending on the particular nature of a sample,
although so far the only numerical work concerns the simple
monopole form, as observed in \cite{mono1}. In the present paper,
the "Snyder" space has yielded an internal magnetic field that is
singular in {\it{coordinate space}} and not in the form of a
monopole.  In fact it depends on the particle magnetic moment.
Also it should be noted that the angular momentum occurs
explicitly in the "Snyder" phase space algebra (\ref{1}). This
indicates the presence of an inherent chiral nature in the phase
space. Hence, we believe that it will be very interesting to study
the response of this particle to an external magnetic field, which
we intend to study in near future. These observations tend to
suggest that  the novel NC model studied here, can serve as an
effective theory for the physical phenomena analyzed in
\cite{gen,pur}. \vskip .3cm {\it{Acknowledgements}}: It is a
pleasure  to thank Banasri Basu and Peter Horvathy for
discussions.


\begin{thebibliography}{99}
\bibitem{sw}N.Seiberg and E.Witten, JHEP 9909(1999)032. For reviews see  for example M.R.Douglas and N.A.Nekrasov,
Rev.Mod.Phys. 73(2001)977; R.J.Szabo, {\it {Quantum Field Theory
on Noncommutative Spaces}}, hep-th/0109162.
\bibitem{sn} H.S.Snyder, Phys.Rev. 71 38(1947).
\bibitem{kappa} J.Lukierski, H.Ruegg,
W.J.Zakrzewski, Annals Phys. 243 (1995)90; S.Majid and H.Ruegg,
Phys.Lett. B334 (1994)348; S.Ghosh and P.Pal, Phys.Lett.B 618
(2005)243, (hep-th/0502192); S.Ghosh, Phys.Lett.B 623 (2005)251,
(hep-th/0506084).
\bibitem{jac1}R.Jackiw, Phys.Rev.Lett. 54(1985)159; Int.J.Mod.Phys. A19S1(2004)137.
\bibitem{cnp}C.Chou, V.P.Nair and A.P.Polychronakos, Phys.Lett.
 B304 105(1993); S.Ghosh, Phys.Lett. B338 (1994) 235 (hep-th/9406089); Erratum-ibid. B347 (1995) 468; Phys.Rev. D51 (1995) 5827 (hep-th/9409169); Erratum-ibid. D52 (1995) 4762.
  \bibitem{wil}For a review, see F.Wilczek, {\it{Fractional Statistics and Anyon Superconductivity}}, World Scientific, Singapore, 1990.
\bibitem{mono1}M.Onoda and N.Nagaosa, J.Phys.Soc.Jpn. 71 19(2002); Z.Fang et al., Science 302 92(2003).
\bibitem{niu}G.Sundaram and Q.Niu, Phys.Rev.B B59(1999)14915;
R.Shindou and K.-I Imura, cond-mat/0411105; C.Zhang, A.M.Dudarev
and Q.Niu, cond-mat/0507125.
\bibitem{spin}J.E.Hirsch, Phys.Rev.Lett. 83 1834(1999); S.Zhang, Phys.Rev.Lett. 85 393(2000); S.Datta and B.Das, Appl.Phys.Lett. 56 665(1990).
\bibitem{berard}A.Berard and H.Mohrbach. Phys.Rev. D69 95 127701(2004)
\bibitem{duv}C. Duval, Z. Horvath, P. A. Horvathy, L. Martina and P. Stichel, Mod.Phys.Lett. B20 (2006) 373-378 (cond-mat/0506051) ; C. Duval, Z. Horvath, P. Horvathy, L. Martina, P. Stichel
 Phys.Rev.Lett. 96 (2006) 099701 (cond-mat/0509806),
Z.Horvath, P.A.Horvathy and L.Martina, {\it{Hamiltonian aspects of
Bogoliubov quasiparticles}},  cond-mat/0511099.
\bibitem{xiao}D.Xiao, J.Shi and Q.Niu, Phys.Rev.Lett. 95 137204(2005).
\bibitem{xiao1}D.Xiao, J.Shi and Q.Niu, Phys.Rev.Lett. 96 099702(2006).
\bibitem{pur}P.Pureur et.al.,  cond-mat/0501482.
\bibitem{gen} G.Tatara and H.Kawamura, J.Phys.Soc.Jpn 71 2613(2002); H.Kawamura, Phys.Rev.Lett. 90 047202(2003); M.Onoda, G.Tatara and N.Nagaosa, J.Phys.Soc. of Japan, 73 2624(2004).
\bibitem{ban}B. Basu, S.Ghosh and S.Dhar, {\it{ Noncommutative Geometry and Geometric Phases}}, hep-th/0604068.
\bibitem{jabb} M.M.Sheikh-Jabbari, Phys.Lett. B455 129(1999) (HEP-TH/9901080);
D.Bigatti and L.Susskind, Phys.Rev. D62 (2000) 066004
(HEP-TH/9908056); S.Ghosh, Phys.Lett. B571 (2003)97.
\bibitem{jack}J.D.Jackson, {\it{Classical Electrodynamics}} (chapter 5), Wiley, 1962.
 \bibitem{fj}L.Faddeev and R.Jackiw, Phys.Rev.Lett. 60 1692(1988).
 \bibitem{dh}C.Duval and P.Horvathy, Phys.Lett.B 479 284(2000).
 \bibitem{chang}L.N.Chang et.al., Phys.Rev. D65 125027(2002) (hep-th/0111181).
 \bibitem{gen1}G.Tatara (private communication).


\end{thebibliography}
\end{document}